\documentclass[aps,pre,twocolumn,showpacs,superscriptaddress,nofootinbib]{revtex4-1}  
\usepackage{graphicx}  
\usepackage{dcolumn}   
\usepackage{bm,amssymb,amsmath}   
\usepackage{xcolor}    
\usepackage[normalem]{ulem}	
\definecolor{mygreen}{rgb}{0.1, 0.6, 0.0}
\begin{document}
\widetext
\title{Interplay between media and social influence\\
in the collective behavior of opinion dynamics}
\author{Francesca Colaiori}
\affiliation{Istituto dei Sistemi Complessi (ISC-CNR), UOS Sapienza, c/o Dipartimento di Fisica, Sapienza Universit\`a di Roma, Piazzale Aldo Moro 5, 00185 Rome, Italy}
\affiliation{Dipartimento di Fisica, Sapienza Universit\`a di Roma, Piazzale Aldo Moro 5, 00185 Rome, Italy}
\author{Claudio Castellano}
\affiliation{Istituto dei Sistemi Complessi (ISC-CNR), Via dei Taurini 19, 00185 Roma, Italy}
\affiliation{Dipartimento di Fisica, Sapienza Universit\`a di Roma, Piazzale Aldo Moro 5, 00185 Rome, Italy}
\date{\today}

\begin{abstract} 
Messages conveyed by media act as a major drive in shaping attitudes and inducing opinion shift.  
On the other hand, individuals are strongly affected by peers' pressure while forming their own judgment.
We solve a general model of opinion dynamics where 
individuals either hold one out of two alternative opinions or are undecided, 
and interact pairwise while exposed to an external influence.  As media
pressure increases, the system moves from pluralism to global consensus; 
four distinct classes of collective behavior emerge, crucially depending on the outcome of
direct interactions among individuals holding opposite opinions.
Observed nontrivial behaviors include hysteretic phenomena and
resilience of minority opinions.  Notably, consensus could be
unachievable even when media and microscopic interactions are biased
in favour of the same opinion: the unfavoured opinion might even gain 
the support of the majority.

\end{abstract}
\maketitle

\section{INTRODUCTION}
Public opinion formation is a complex mechanism: people constantly
process a huge amount of information to make their own judgment.
Individuals form, reconsider and possibly change their opinions under
a steady flow of news and a constant interchange with others.  Social
interactions have a strong effect on opinion
formation~\cite{Buchanan2008}: the pressure to conform to the opinion
of others may in some cases even overcome empirical
evidence~\cite{Asch1955} or undermine the wisdom of the
crowd~\cite{Lorenz2011}.  On the other hand media also play a central
role: they can influence public knowledge, attitudes and behavior by
choosing the slant of a particular news story, or just by selecting
what to report.  An intense activity has recently investigated
macroscopic effects of these fundamental mechanisms exploiting the
connection between opinion dynamics and simple non--equilibrium
statistical physics models~\cite{Castellano2009,Sen2013}.  Media
influence has been considered in the literature, in particular within
the framework of cultural dynamics~\cite{Gonzalez-Avella2005,
  Gonzalez-Avella2010, Peres2011} and continuous opinion dynamics with
heterogeneous confidence bounds~\cite{Carletti2006, VazMartins2010,
  Quattrociocchi2014, Pineda2015}.  The fundamental case of binary
opinions has received less attention for what concerns the role of
media~\cite{Bordogna2007, Crokidakis2011, Michard2005}, although the
similar case of proselytism by committed agents (zealots) has
attracted considerable interest~\cite{Mobilia2003,Xie2011}.  The case
of binary opinions is clearly relevant to yes/no questions, but not
limited to that: when people are prompted with important questions
that admit many possible answers their attitudes tend to be polarized,
most people sharing one out of two opposite opinions
\cite{Vallacher1994}.

In this paper we address the problem of how people form their opinions
based on the message they receive, both from traditional media and
their social network, by studying a general class of opinion dynamics
models.  At any given time, each individual in the population is
either supporting one of two alternative opinions or ``undecided". The
possibility to be in a third state may have crucial effects in the
case of a binary choice~\cite{Vazquez2003, DeLaLama2006, Castello2006,
  Colaiori2015}.  Individuals are exposed to some external source of
information biased towards one opinion (mainstream) and exchange
information upon pairwise interaction: both factors may cause them to
change their state.  The undecided state accounts for individuals
being uninterested, uninformed, or generally confused on the given
issue.  Carrying no opinion, they are assumed to have a passive role
in the interactions.  We account for the effect of media in the
simplest possible way by assuming that people have some general
tendency to conform to the media recommendation.  On the other hand,
we consider totally general rules for the interactions among pairs of
agents. Our general model encompasses therefore a large class of
specific models, each one identified by a given set of parameters
defining the interaction rules. 
We study in mean-field this general model, determining
the stationary solutions and their stability as a function of
the external bias and of the parameters specifying pair interactions.
We uncover the emergence of four distinct classes of collective
behavior, characterized by different responses to the media exposure.
Only two linear combinations of the parameters defining the general
dynamics are relevant in determining which class a specific model belongs to.
The results of numerical simulations, performed in systems with 
interaction patterns described by complex networks,
support the general validity of the MF picture.

\section{GENERAL MODEL}
We now define in detail the general model.  Each agent can be in one
of three states: holding opinion $A$, holding an opposing opinion $B$,
or being undecided ($U$). Agents tend to conform to the media
recommendation by adopting opinion $A$ at a constant rate $r$,
independently of their current state, and interact pairwise at rate
$f$.  In the following we set $f=1$ to fix the time scale. Individuals
in the same state are unaltered by their mutual interaction.  In
$A-U$($B-U$) interactions undecided agents may adopt, with given probability, their partner's
opinion, that they cannot alter: an agent
holding opinion $A$ ($B$) has a probability $\varphi_2$ ($\gamma_2$)
to convince the $U$ agent.  We assume in general $\varphi_2 \ne
\gamma_2$, allowing for $A$ and $B$ to have unequal efficacy in
persuading others.  Interactions among agents holding opposite
opinions ($A-B$) may have any outcome: each of the two agents may
either keep her opinion, change it to match her partner's opinion, or
get confused and turn to the undecided state, in any combination.  We
indicate with $\left\{\alpha_i\right\}_{i=1}^6$ the probabilities of
the six possible outcomes.  Seven
independent parameters fix the probabilities of each possible outcome
for any interacting pair, as summarized in Table~\ref{interactions}.
While we consider for simplicity symmetric roles for the two
interacting partners, our mean--field (MF) analysis holds more
generally, also including models that assign distinct roles
(e.g. speaker/listener) to the two partners. This more general case is
discussed in Appendix A.  Note that asymmetric models
are always equivalent, in mean field, to their symmetrized version,
the outcome of a symmetric interaction being the average result of
asymmetric interactions with exchanged roles.
\begin{table}
\begin{ruledtabular}
\begin{tabular}{cccccccc}      
     &&$A,A$     & $B,B$    & $U,U$    & $A,U$      & $B,U$      & $A,B$    \\ 
\hline
$A-A$&&$1$       &$0$       &$0$       &$0$         &$0$         & $0$      \\
$B-B$&&$0$       &$1$       &$0$       &$0$         &$0$         & $0$      \\
$U-U$&&$0$       &$0$       &$1$       &$0$         &$0$         & $0$      \\
$A-U$&&$\varphi_2$&$0$      &$0$       &$1-\varphi_2$&$0$        & $0$      \\
$B-U$&&$0$       &$\gamma_2$&$0$       &$0$         &$1-\gamma_2$& $0$      \\
$A-B$&&$\alpha_1$&$\alpha_2$&$\alpha_3$&$\alpha_4$   &$\alpha_5$  &$\alpha_6$\\
\end{tabular}
\caption{Each row in the table corresponds to an interaction, and each
  column to a possible outcome. Elements in the table indicate the
  probabilities of each possible outcome for the given interaction. In
  the last row $\sum_{i=1}^6\alpha_i=1$.}
\label{interactions}
\end{ruledtabular}
\end{table}

\section{MEAN FIELD EQUATIONS}
The evolution of the system is described in MF by the dynamical equations for the density of agents in each state:
\begin{equation}
\!\!\left\{
\begin{array}{lll} 
\!\! \dot{n}_A&
\!\!=\!\!\!\!&
r(1-n_A)+2\varphi_1 n_A n_B +2 \varphi_2 n_A (1-n_A-n_B)\!\!\!\!\\
\!\!\dot{n}_B &
\!\!=\!\!\!\!&
-r n_B+2\gamma_1 n_A n_B  +2 \gamma_2 n_B (1-n_A -n_B )  
\end{array}
\!\!\!\right.
\label{dynamics}
\end{equation}
where $n_A$ ($n_B$) denote the density of agents in the $A$ ($B$)
state, and $\varphi_1=\alpha_1-\alpha_2-\alpha_3-\alpha_5$,
$\gamma_1=-\alpha_1+\alpha_2-\alpha_3-\alpha_4$.  The density $n_U$ of
undecided agents is obtained by normalization: $n_U=1-n_A-n_B$.  The
densities must belong to the {\it physical region} of the plane
$(n_A,n_B)$, defined by the three constraints $n_A\geq 0$, $n_B\geq
0$, $n_A+n_B \leq 1$.  The coefficients $\alpha_i$ appear in
Eq.~(\ref{dynamics}) only in two linear combinations $\varphi_1$ and
$\gamma_1$ that represent the total average variation in $A$ ($B$)
states due to an $A-B$ interaction.  This reduces the number of
effective parameters from seven to four plus the external bias $r$,
acting as a control parameter.  The parameters $\varphi_1$ and
$\gamma_1$ are bounded by $-1\leq \gamma_1 \leq 1$, $-1 \leq
\varphi_1\leq1$, $\gamma_1+\varphi_1\leq 0$ (the sum
$\gamma_1+\varphi_1$ is minus the average net production of undecided
in an $A-B$ interaction, that has to be non negative).  Stationary
solutions of Eq.~(\ref{dynamics}) are the intersections of the two
conic sections:
\begin{equation}
\!\left\{
\begin{array}{lr}
\!\!n_A\left[(\varphi_2  -\varphi_1  )n_B + \varphi_2   n_A-\varphi_2  +r/2\right]=r/2 
& \,\,\,{\cal C}_1\!\!\\
\!\!n_B\left[(\gamma_1-\gamma_2) n_A - \gamma_2 n_B+\gamma_2-r/2\right]=0  
& \,\,\, {\cal C}_2\!\!
\end{array}
\!\!\right.
\label{curves}
\end{equation}
The curve ${\cal C}_1$ is an hyperbola. One of its asymptotes is the
axis $n_A=0$, so that only the upper branch ${\cal C}_1^{+}$ ($n_A>0$)
is physically relevant.  The curve ${\cal C}_2$ factorizes into the
product of two lines, ${\cal R}_1$, and ${\cal R}_2$.  The line ${\cal
  R}_1$ ($n_B=0$) always intercepts ${\cal C}_1^{+}$ in $P_1 \equiv
[n_B=0,n_A=1]$, corresponding to the absorbing state of total
consensus on opinion $A$.  Depending on the parameters, ${\cal R}_2$
may intersect ${\cal C}_1^{+}$ in one, two (possibly coincident)
points, either inside or outside the physical region, or none.
Varying the control parameter $r$, different fixed points arise and
move entering and exiting the physical region.  Their flow and
stability determine the collective response to the external bias.
\begin{figure}
\centering
\includegraphics[width=8.5cm]{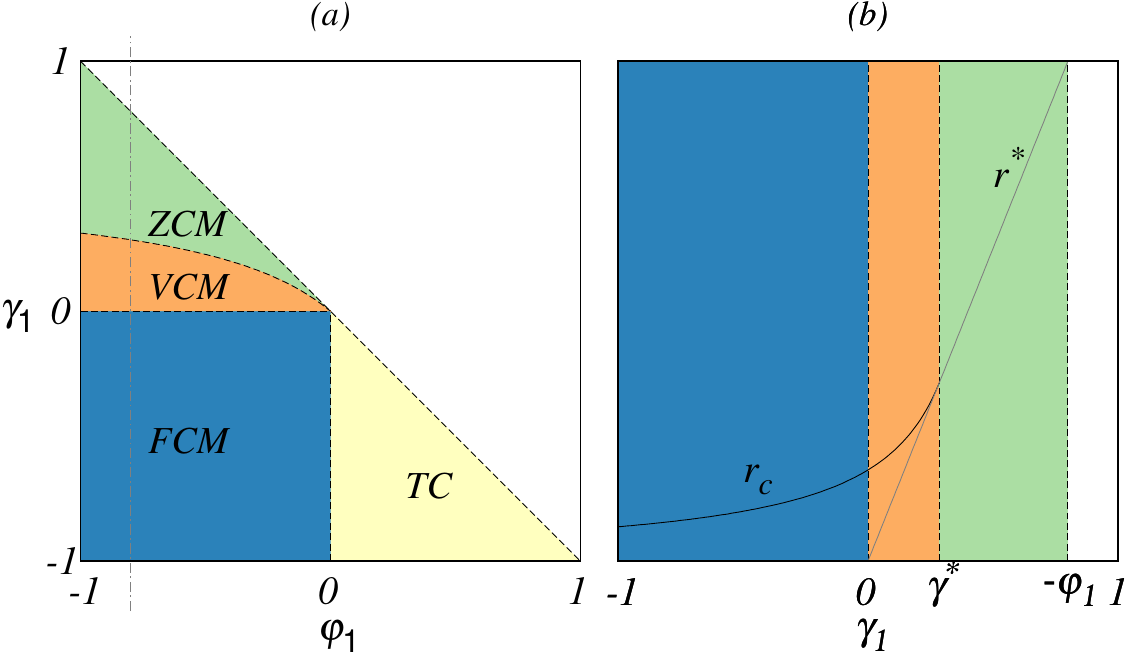}
\caption{(color online)(a) Phase diagram in the ($\varphi_1$,
  $\gamma_1$) plane.  The curve $\gamma_1^{*}$ separating regions
  $VCM$ and $ZCM$ depends on $\varphi_2$ and $\gamma_2$ (in the figure
  $\varphi_2=0.1$ and $\gamma_2=0.5$).  (b) Plot of $r_c$
  (saddle--node bifurcation curve) and $r^*$ (transcritical
  bifurcation line) as functions of $\gamma_1$, and for
  $\varphi_1=-0.8$. }
\label{fig2:PhaseDiagram}
\end{figure}
By studying them we find the phase-diagram represented in 
Fig.~\ref{fig2:PhaseDiagram}(a), which constitutes the main result of
our paper. As a function of the external bias $r$ there are
four distinct classes of collective behavior, associated with 
different regions of the parameter space. Notably, the emergent behavior 
is essentially ruled by only two of the parameters, $\varphi_1$ and $\gamma_1$.
In the next section we discuss the qualitative features of each class, 
referring to Appendix B for analytical details on the derivation of the phase
diagram.
\begin{figure*}
\centering
\includegraphics[width=\textwidth]{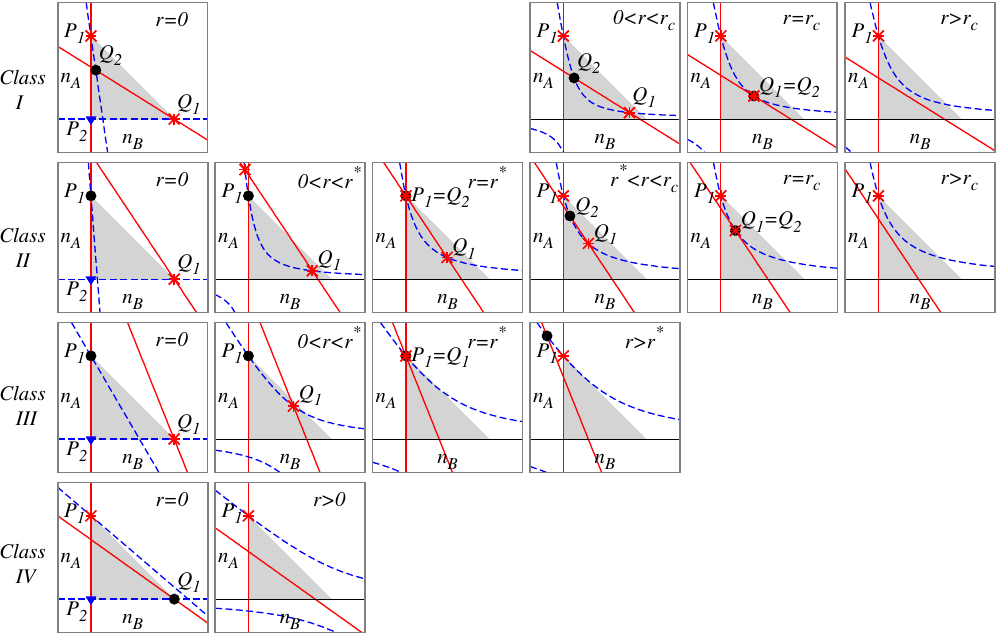}
\caption{Plots of isoclines and fixed points for each class of models
  and each range of $r$. Red solid lines are ${\cal R}_1$ and ${\cal
    R}_2$, the blu dashed curve is the hyperbola ${\cal C}_1$. Red
  stars denote stable fixed points, black circles denote saddle
  points, blue triangles denote repulsive fixed points.  The shaded
  triangle represents the physical region $n_A\geq 0$, $n_B\geq 0$,
  $n_A+n_B\leq1 $. Plots in the first row correspond to
  $\varphi_1=-0.6$, $\varphi_2=0.1$, $\gamma_1=-0.3$, $\gamma_2=0.5$,
  and $r=0$, $r=r_c/2$, $r=r_c$, $r=r_c+0.05$. Plots in the second row
  correspond to $\varphi_1=-0.6$, $\varphi_2=0.05$, $\gamma_1=0.1$,
  $\gamma_2=0.3$, and $r=0$, $r=r^*/2=\gamma_1$, $r=r^*=\gamma_1$,
  $r=0.2 r^*+0.8 r_c$, $r=r_c$, $r=r_c+0.05$. Plots in the third row
  correspond to $\varphi_1=-0.5$, $\varphi_2=0.7$, $\gamma_1=0.3$,
  $\gamma_2=0.5$, and $r=0$, $r=r^*/2=\gamma_1$, $r=r^*=\gamma_1$,
  $r=r^*+0.1$. Plots in the fourth row correspond to $\varphi_1=0.1$,
  $\varphi_2=0.7$, $\gamma_1=-0.2$, $\gamma_2=0.5$, and $r=0$,
  $r=0.2$.}
\label{PD}
\end{figure*}
\begin{figure*}
\centering
\includegraphics[height=3.8cm]{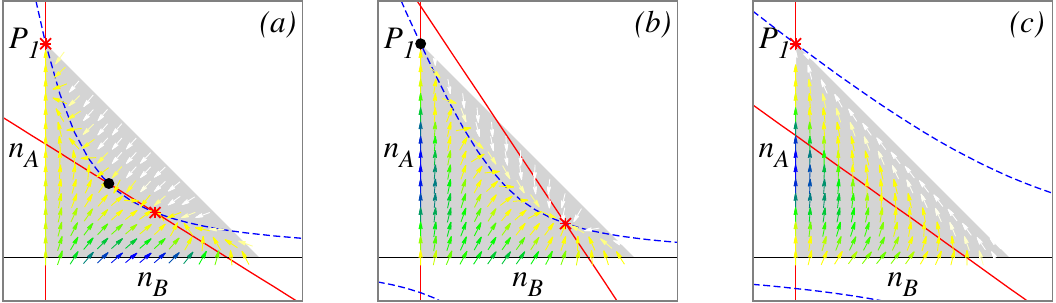}
\caption{(color online) Phase portrait relative to Eq.~(\ref{dynamics}):
  flows in the vector field indicate the time evolution of the system.
  Fixed points are the intersections between either of the two lines
  ${\cal R}_1$, ${\cal R}_2$ (red solid lines), and the curve ${\cal
    C}_1$ (blue dashed line).  Black circles are saddle points, red
  stars are attractive fixed points.  (a) $FCM$ model in the region
  $r<r_c$ ($\varphi_1=-0.6$, $\varphi_2=0.1$, $\gamma_1=-0.3$,
  $\gamma_2=0.5$, $r=0.1$); (b) $VCM$ model in the range $r<r^*$
  ($\varphi_1=-0.6, \varphi_2=0.4$, $\gamma_1=0.1$, $\gamma_2=0.3$,
  $r=0.13$); (c) $TC$ model ($\varphi_1=0.1$, $\varphi_2=0.7$,
  $\gamma_1=-0.2$, $\gamma_2=0.5$, $r=0.2$).  The flow is
  qualitatively similar to (a) for $VCM$ models with $r^*<r<r_c$, to
  (b) for $ZCM$ models with $r<r_c$, to (c) for any class of models
  above $r_c$. See Appendix B for further details. }\label{Directionfield}
\end{figure*}
\begin{figure*}
\centering
\includegraphics[width=\textwidth]{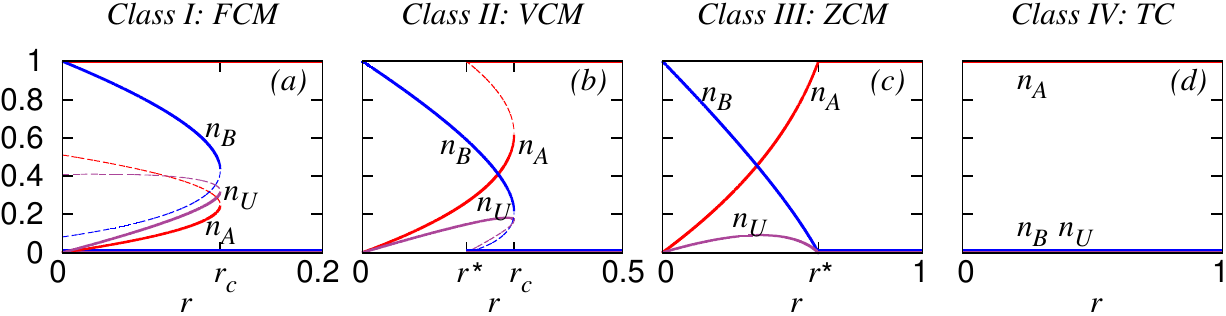}
\caption{(color online) Theoretical results for the densities of
  agents for realizations of each of the four classes of models: (a)
  {\em Class I} model (FCM) ($\varphi_1=-0.5$, $\varphi_2=0.1$,
  $\gamma_1=-0.4$, and $\gamma_2=0.5$), (b){\em Class II} model (VCM)
  ($\varphi_1=-0.5$, $\varphi_2=0.1$, $\gamma_1=0.1$, and
  $\gamma_2=0.5$), (c){\em Class III} model (ZCM) ($\varphi_1=-0.5$,
  $\varphi_2=0.1$, $\gamma_1=0.3$, and $\gamma_2=0.5$), (d){\em Class
    IV} model (TC) ($\varphi_1=0.3$, $\varphi_2=0.1$, $\gamma_1=-0.5$,
  and $\gamma_2=0.5$).  Solid (dashed) lines represent stable
  (unstable) lines. }
\label{Densities}
\end{figure*}
Fig.~\ref{PD} represents the shapes of the curves in Eq.~(\ref{curves}),
allowing to understand the existence and positions of stationary solutions.
Fig.~\ref{Directionfield} provides information on flows and the stability
of solutions.
Fig.~\ref{Densities} depicts the resulting behavior of the
densities of agents in the different states as a function of $r$.

\section{CLASSES OF COLLECTIVE BEHAVIOR}

\subsection{Class I: Finite Critical Mass (FCM) Models} 
When $\varphi_1<0$, $\gamma_1 \leq 0$, i.e. in models where $A-B$
interactions produce on average
an increase in undecided individuals with no net gain in $A$ nor in $B$ states, 
the system undergoes a first
order transition at a finite value $r=r_c$ of the external bias
(see Fig.~\ref{PD}, first row).
For large enough $r$ $P_1$ is the only fixed point and the system flows 
into the absorbing state of total consensus on opinion $A$ for 
any initial condition.  At $r=r_c$ the
system undergoes a saddle--node bifurcation \cite{Strogatz2001}: two
additional coinciding fixed points appear. As $r$ is decreased below
$r_c$ they split (one, with larger
$n_B$, stable, the other unstable, see Fig.~\ref{Directionfield}(a)).  
In the nontrivial stable fixed point the two opinions $A$ and $B$ coexist 
in the population, together with a fraction of undecided (we call
this state ``pluralism"), see Fig.~\ref{Densities}(a).  The initial
conditions determine whether the pluralistic state ($n_B>0$) or the
consensus state ($n_A=1$) is asymptotically reached.  The value of
$n_B$ at the unstable fixed point stays finite in the limit $r
\rightarrow 0$ (see Fig.~\ref{PD}, first row), 
implying that a finite ``critical mass" \cite{Dodds2004} of dissenters 
is always needed to reach the pluralistic state, no matter how small 
is the external bias.

\subsection{Class II: Vanishing Critical Mass (VCM) Models}
This class is identified by $\varphi_1<0$, $0< \gamma_1 < \gamma_1^*$,
and corresponds to models where $A-B$ interactions cause on average a 
small increase of $B$ states at the expense of $A$ states.  
As in {\em Class I}, lowering $r$ below $r_c$
the system undergoes a first order transition separating a regime
($r>r_c$), where consensus on opinion $A$ is the only stable state
from a regime ($r<r_c$) where a stable
and an unstable additional fixed points appear through a saddle--node
bifurcation (Fig.~\ref{PD}, second row).  However, in this case,
further reducing $r$, the unstable fixed point collides with the point
$P_1$ (consensus on $A$) at a finite value $r^*$ ($0<r^*<r_c$), and
then exits the physical region.  This
is a transcritical bifurcation \cite{Strogatz2001}: when the two fixed
points cross each other, they exchange stability 
(Fig.~\ref{Directionfield}(b)); 
$P_1$ becomes unstable, so that below $r^*$ the system, unless started with 
$n_B \equiv 0$, always flows to the pluralistic state
(Fig.~\ref{Densities}(b)).  The initial presence of even a few
dissenters suffices for opinion $B$ to survive. 
The curve
$\gamma_1^*(\varphi_1) = \varphi_1\gamma_2/(\varphi_1-\varphi_2-\gamma_2)$
separating {\it Class II} and {\it III} depends on the parameters
$\varphi_2$ and $\gamma_2$ (see Appendix B).  
The transcritical bifurcation line is
$r=r^*=2\gamma_1$ (see Appendix B), and always lies below the saddle--node
bifurcation line (see Fig.~\ref{fig2:PhaseDiagram}(b)).

\subsection{Class III: Zero Critical Mass (ZCM) Models}
When $\varphi_1<0$, $\gamma_1 \geq \gamma_1^*$, corresponding to
models where $A-B$ interactions give an increase of undecided and 
a large increase in $B$ at the
expense of $A$, the system undergoes at $r=r^*=2 \gamma_1$ a
continuous transition (transcritical bifurcation) between total
consensus on opinion $A$ ($r>r^*$) and pluralism ($r<r^*$), see
Fig.~\ref{PD}, third row, and Fig.~\ref{Densities}(c).  
Initial conditions do not play any role.

\subsection{Class IV: Total Consensus (TC) Models}
The region $\varphi_1\geq 0$ corresponds to models where $A-B$
interactions result in a net increase of individuals holding opinion
$A$.  The behavior is trivial (see Fig.~\ref{Directionfield}, fourth row): 
irrespectively of the value of all other parameters, for any initial 
condition, and no matter how small the external forcing is, the system 
always converges to the consensus state ($P_1$) 
(see Fig.~\ref{Directionfield}(c) and~\ref{Densities}(d)).


\section{BEYOND MEAN--FIELD}

Before discussing the analytical results obtained in the previous
section within MF and examining some of their consequences we test
their validity beyond MF by numerical simulations on synthetic and
real--word networks. The simulations are performed on four microscopic
dynamical models, each belonging to one of the universality classes
derived above.

The single event of the dynamics occurs as follows~\cite{nota2}.  
We select randomly a node $i$ (speaker) and, with probability $r/(1+r)$, we set
his state to $A$, as effect of the external bias.  Instead, with
complementary probability $1/(1+r)$ an interaction process takes
place: we select a listener $j$ among the neighbors of $i$, and modify
the state of the pair ($i$, $j$) according to Table~\ref{interactions2}
in Appendix B with probabilities $\psi_2=\omega_2=\varphi_2$,
$\delta_2=\epsilon_2=\gamma_2$ and each $\lambda_i=\mu_i=\alpha_i$ for
any $i$.  Each single event occurs during a temporal interval $1/N$
(where $N$ is the total number of nodes in the network), so that $N$
updates are attempted in a time unit.  Starting from the initial
configuration, for each value of $r$ we let the system evolve during
5000 time steps to reach the stationary configuration and determine
the densities $n_A$, $n_B$ and $n_U$ by performing averages over 5000
additional time steps.  In order to characterize the possible presence
of discontinuous phase transitions and the associated hystereric
effects, for each set of data we consider two different initial
conditions: either all nodes in state $B$ ($n_B=1$), or all nodes in
state A except for a very small fraction of nodes in state $B$
($n_A=0.99$, $n_B=0.01$).  We always keep the values $\varphi_2=0.1$
and $\gamma_2=0.5$ fixed and vary $\gamma_1$ and $\varphi_1$ in order
to encompass all four quantitatively distinct behaviors found in the
analytic approach.
\begin{figure*}
\centering\
\includegraphics[width=\textwidth]{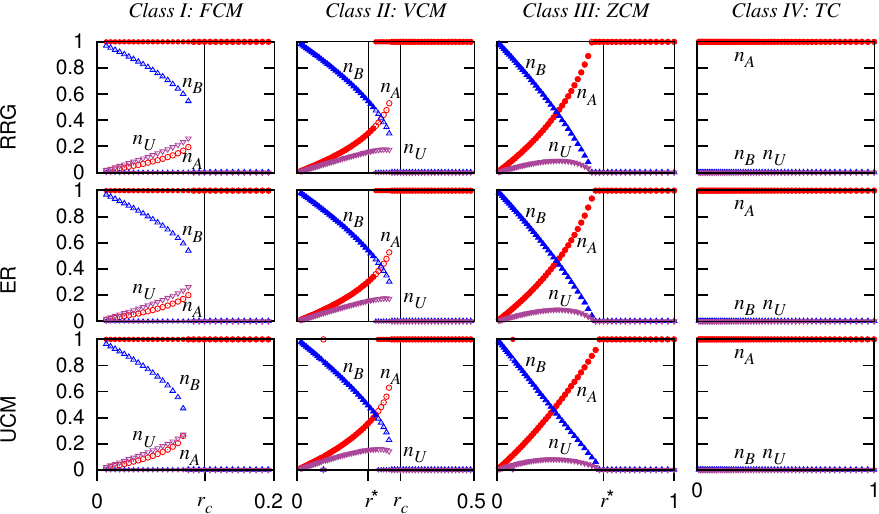}
\vspace{.05cm}
\includegraphics[width=\textwidth]{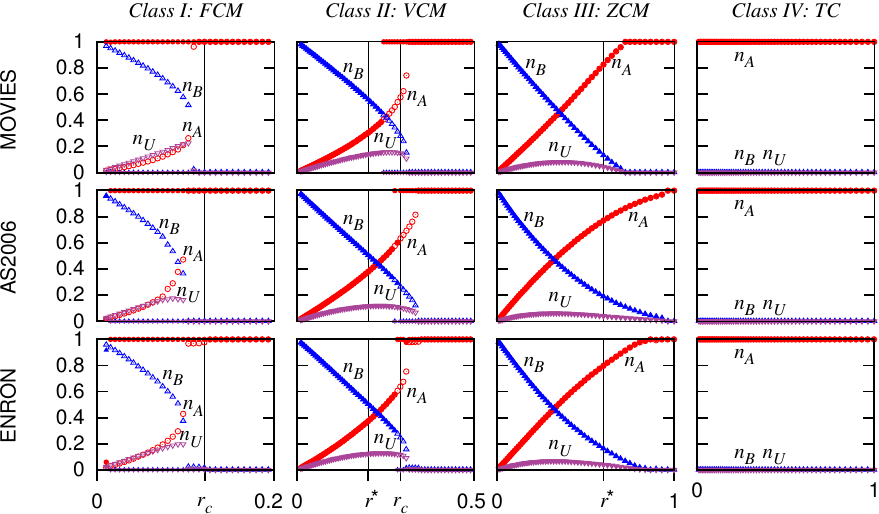} \caption{(color online)  Stationary densities of nodes in
  the various states as a function of the external bias $r$ for three
  types of synthetic networks and three real networks described in the text. Empty (filled) symbols
  refer to the $n_B=1$, $n_A=0$ ($n_B=0.01$, $n_A=0.99$) initial condition.  
  Red circles, blue upward triangles, violet
  downward triangles represent $n_A$, $n_B$, and $n_U$, respectively.  
  The values of $\varphi_2=0.1$
  and $\gamma_2=0.5$ are fixed.  From left to right the values of the
  pair $(\varphi_1,\gamma_1)$ are $(-0.5,-0.4)$, $(-0.5,0.1)$,
  $(-0.5,0.3)$ and $(0.1,-0.5)$, corresponding respectively to classes
  {\em I}, {\em II}, {\em III} and {\em IV} according to the MF classification scheme. Notice that in class {\em I}, $r^* \approx 1$ for AS2006, but since $r$ is a rate   there is still an extended consensus region for $r>r^*$.}
   \label{RRG}
\end{figure*}

\subsection{Simulations on synthetic networks}

In Fig.~\ref{RRG} (upper panel) 
we plot the stationary value of the densities as a
function of $r$ when the interaction pattern is given by three
synthetic networks of size $N=20000$: a Random Regular Graph (RRG)
where each node has 10 neighbors; an Erd\"os-R\'enyi(ER) graph of 
average degree 10; a network built using the
Uncorrelated Configuration Model (UCM) with minimum degree $3$
and degree distribution decaying as $k^{-2.5}$.
For all these cases, predictions
are very well matched by numerical simulations.

\subsection{Simulations on real networks}

A tougher test of the MF results is provided by simulations
performed on real--world networks, incorporating additional
topological features such as clustering and correlations.  In
Fig.~\ref{RRG} (lower panel) we report results for: a network of size $N=81860$
representing movie actor collaborations obtained from the Internet
Movie Database (MOVIES) (average degree $\langle k \rangle=89.53$, 
fluctuations $\langle k^2 \rangle/\langle k \rangle = 594.91$); 
a network of size $N=24608$ representing
connections of Internet Autonomous Systems in 2006 (AS2006) 
(average degree $\langle k \rangle=4.05$, 
fluctuations $\langle k^2 \rangle/\langle k \rangle =  259.94$) ; 
the largest connected component (size $N=33696$) of the Enron email
exchange network (ENRON) (average degree $\langle k \rangle=10.02$, 
fluctuations $\langle k^2 \rangle/\langle k \rangle =  140.07$)~\cite{snapnets}.

Also in these cases numerical simulations agree well with the outcome
of the MF approach: the behavior in each of the classes
qualitatively reproduces the analytical predictions, with only (expected)
variations in the position of transition points.  The only
variation in this respect concerns class I. In this case it turns out
that for very small values of the rate $r$ the state with overwhelming
majority of $B$ is dynamically reached even starting from $n_B$ as low
as $0.01$, at odds with the MF prediction.

\section{DISCUSSION AND CONCLUSIONS}

We finally discuss some interesting and nontrivial consequences that follow from the 
classification scheme derived within the MF approach.  

Although the parameters
regulating the interactions with undecided agents have a marginal role
in determining the collective behavior, the presence itself of the $U$
state is crucial in several respects.  In the absence of the third,
undecided state, the system would either converge
to total consensus on opinion $A$ ({\it Class IV})-- when asymmetric
interactions favor $A$, or exhibit a continuous transition between
consensus and pluralism ({\it Class III}) -- when asymmetric
interactions favor $B$~\cite{Colaiori2015}.

Several models in the literature that allow for a third state require
$U$ to be a necessary intermediate step when changing
opinion~\cite{Marvel2012, Castello2006}.  Our results imply that such
systems undergo a discontinuous transition by varying the media
exposure: $U$ being a necessary intermediate step requires
$\alpha_1=\alpha_2=0$, giving $\varphi_1 \leq 0$, $\gamma_1 \leq 0$;
therefore these models always fall in {\em Class I}.  The condition
for a model to be in {\it Class I} is however more general, only
excluding {\it average} gain of $A$ or $B$ in $A-B$ interactions.

Within our framework consensus is always achieved, whatever the
interactions, for strong enough media exposure.  Then a natural
question arises: is consensus stable upon removal of the media
pressure?  The answer to this question is different for different
classes.  In FCM class, once the consensus is reached, it is kept also
when the media exposure is removed.  In ZCM class dissenters nucleate
as soon as the media exposure is lowered below the threshold needed to
reach consensus ($r<r_c$).  VCM class shows interesting hysteretic
behavior: once consensus is reached with a sufficiently high media
exposure ($r>r_c$), it is kept when lowering the media pressure that
however cannot be completely turned off: below $r^*$ consensus on opinion $A$
becomes unstable and any infinitesimal perturbation causes an abrupt
transition to a pluralistic state.

Finally, a markedly counter--intuitive fact emerges from our analysis,
namely the possibility for $B$ states to survive, and even become the
majority, also when both the external forcing and the interaction
rules are biased against them.  Examples are systems in the FCM class that have
$\gamma_1<\varphi_1$ (more $B$ than $A$ states are lost in $A-B$
interactions), and $\varphi_2>\gamma_2$ (individuals in $A$ state have
more success than those in $B$ state in convincing undecided
individuals).  In this case, both the external pressure and the rules
of peer interactions favor $A$, yet for small values of $r$ and
suitable initial conditions a stationary state with $n_B>0$ is
reached.  Here the role of the third state is crucial: although the
peer interaction rules are asymmetric in favor of the $A$ state, the
rate at which they occur allows the $B$ state to be favored on average~\cite{nota1}.

Analytical predictions derived in MF are well matched by numerical
simulations performed both on synthetic and real networks: the
collective behavior for sample models in each of the four classes
qualitatively reproduces the analytical predictions.  The only
observed deviation from MF concerns models in class I on some real
topologies.  In this case, for very small values of the rate $r$, even
when the initial state has $n_B$ as low as $0.01$, the state with
overwhelming majority of $B$ is dynamically reached while MF would
predict that total consensus is achieved for suitably small values of
$n_B$ in the initial condition. This could reflect
topological features such as clustering or correlations present
in the networks.


In conclusion, we have shown how four qualitatively distinct kinds of
collective dynamics emerge from the interplay between mass media and
social influence, and categorized a very extensive set of opinion
dynamics models accordingly.  The four classes are non--degenerate in
the sense that they occupy a finite region of the parameter space.
The macroscopic behavior is independent of many details and
essentially determined by the outcome of direct interactions among
agents holding opposite opinions.  While the existence of undecided
individuals is crucial, the parameters that define their interactions
only participate in locating the fixed points and the line separating
classes {\it II} and {\it III}.  Nontrivial effects, including
dependence on initial conditions and history--dependence are observed.

The existence of a pluralistic state is desirable in many
circumstances, but it could lead to a deadlock when unanimous
agreement is required.  Assuming very general interactions among peers
we give conditions for a pluralistic state to exist and survive an
external pressure under the schematic hypothesis that people generally
tend to conform to the message conveyed by the media.  An interesting
direction for further research would include within our framework more
sophisticated descriptions of how public opinion is shaped by media,
such as the ``two--step flow" model~\cite{Katz1955, Lazarsfeld1968}
accounting for the role of opinion leaders or as in more recent
theories~\cite{Dodds2007} focusing on information cascades triggered
by a critical mass of easily influenced individuals.

\appendix
\section*{Appendix A: Generalization to models with asymmetric roles}

In the main text we considered generic rules for the peer
interactions, but we assumed symmetric roles for the two interacting
partners.  However we mentioned that our mean--field analysis holds
more generally, including cases where the interaction partners have
distinct roles (e.g. speaker/listener), as often considered in the
literature.  We here introduce a further generalization of our model
that allows for asymmetric roles, for which all the results derived in
mean--field still hold.  This consists in replacing the peer
interactions with those in Table~\ref{interactions2}.
\begin{table}[h]
\begin{ruledtabular}
\begin{tabular}{ccccccccccc}      
 S $\,\,\,\,\,$L   &&$\,AA\,$     & $\,BB\,$    & $\,UU\,$    & $AU$      & $UA$      & $BU$ & $UB$      & $\,AB\,$      & $\,BA\,$   \\ 
\hline
$A-A$&&$1$       &$0$       &$0$       &$0$         &$0$         & $0$  & $0$  & $0$  & $0$\\
$B-B$&&$0$       &$1$       &$0$       &$0$         &$0$         & $0$  & $0$  & $0$  & $0$\\
$U-U$&&$0$       &$0$       &$1$       &$0$         &$0$         & $0$   & $0$  & $0$  & $0$\\
$A-U$&&$\psi_2$&$0$      &$0$       &$1\!-\!\psi_2$&$0$    & $0$   & $0$  & $0$  & $0$\\
$U-A$&&$\omega_2$ &$0$       &$0$       &$0$          &$1\!-\!\omega_2$  &$0$    & $0$  & $0$  & $0$\\
$B-U$&&$0$       &$\delta_2$&$0$       &$0$          & $0$ &$1\!-\!\delta_2$& $0$   & $0$  & $0$\\
$U-B$&&$0$       &$\epsilon_2$&$0$       &$0$         & $0$  & $0$  &$1\!-\!\epsilon_2$& $0$    & $0$\\
$A-B$&&$\lambda_1$&$\lambda_2$&$\lambda_3$&$\lambda_4$   &$\lambda_5$  &$\lambda_6$&$\lambda_7$&$\lambda_8$&$\lambda_9$\\
$B-A$&&$\mu_1$&$\mu_2$&$\mu_3$&$\mu_4$   &$\mu_5$  &$\mu_6$&$\mu_7$&$\mu_8$&$\mu_9$\\
\end{tabular}
\caption{Each row in the table corresponds to an interaction between a
  speaker, leading the conversation, and a listener, and each column
  to a possible outcome, given by an ordered couple
  (speaker,listener). Elements in the table indicate the probabilities
  of each possible outcome for the given interaction. In the last two
  rows $\sum_{i=1}^9\lambda_i=1$, and $\sum_{i=1}^9\mu_i=1$.}
\label{interactions2}
\end{ruledtabular}
\end{table}

The very complicated system defined by the interactions in
Table~\ref{interactions2} is still described in mean--field by
Eqs.~(\ref{dynamics}) once we properly define the coefficients in
Eqs.~(\ref{dynamics}) as functions of the 22 parameters in
Table~\ref{interactions2}.  In particular, we have
$\varphi_2=(\psi_2+\omega_2)/2$, $\gamma_2=(\delta_2+\epsilon_2)/2$,
$\varphi_1=(\lambda_1+\mu_1-\lambda_2-
\mu_2-\lambda_3-\mu_3-\lambda_5-\mu_5)/2$, and
$\gamma_1=(-\lambda_1-\mu_1+\lambda_2+\mu_2-
\lambda_3-\mu_3-\lambda_4-\mu_4)/2$.  We could also allow agents to
exchange their state in an $A-U$ interaction giving $UA$ as outcome
(and similarly for $U-A$, $B-U$, and $U-B$ interactions): our analysis
and results would still hold, however this goes beyond our
interpretation of agents in state $U$ as having no opinion or information
to convey.

We note that the irrelevance of the distinction between roles holds in 
mean--field in full generality: a model allowing for asymmetric roles is
always equivalent at the mean--field level to its symmetrized version,
the outcome of an interaction in the symmetrized model being defined
as the average result of two asymmetric interactions with exchanged
roles.

\section*{Appendix B: Analytical details of the mean--field analysis}
We present here a detailed derivation of the phase diagram discussed
in the main text. 
The general model is described by Eqs.~(\ref{dynamics})
for the densities of agents in $A$ and $B$ states, the density $n_U$
of undecided agents being $n_U=1-n_A-n_B$. 
Stationary solutions are given by the intersections of the two conic 
sections ${\cal C}_1$ and ${\cal C}_2$ in Eqs.~(\ref{curves}).

Excluding the cases $\varphi_2=0$ and $\gamma_2=0$ that will be
discussed separately, and defining
$\varphi=(\varphi_1-\varphi_2)/\varphi_2$ and
$\gamma=(\gamma_1-\gamma_2)/\gamma_2$ we rewrite Eqs.~(\ref{curves})
as:
\begin{equation}
\!\left\{\begin{array}{lr}
\!\!n_A\left(n_A-\varphi n_B+r/(2\varphi_2)-1\right)=r/(2\varphi_2) 
& \,\,\,{\cal C}_1\!\!\\
\!\!n_B\left(\gamma n_A - n_B+1-r/(2\gamma_2)\right)=0  
& \,\,\, {\cal C}_2\!\!.
\end{array}
\!\!\right.
\label{curves2}
\end{equation}
Since physically relevant solutions must belong to the domain $n_A\geq
0$, $n_B\geq 0$, $n_A+n_B\leq 1$, only the upper branch ${\cal C}_1^+$
of the hyperbola ${\cal C}_1$ matters:
\begin{equation}
n_A=\left(\varphi n_B  +1-r/(2\varphi_2)+\sqrt{\Delta_1}\right)/2 \,\,\,\,\,\, {\cal C}_1^{+}\!\!,
\label{ub}
\end{equation}
where $\Delta_1=(\varphi n_B  +1-r/(2\varphi_2))^2+2r/\varphi_2$. The conic section ${\cal C}_2$ is degenerate and factorizes in the two lines:
\begin{equation}
\begin{array}{lr}
n_B=0 & \,\,{\cal R}_1 \\
n_A=n_B/\gamma+
(r/(2\gamma_2)-1)/\gamma& \,\,{\cal R}_2 \!.
\end{array}
\label{lines}
\end{equation}

${\cal C}_1$ and ${\cal R}_1$ always have two intersections:
$P_1=[n_B=0,n_A=1]$, corresponding to total consensus on opinion $A$,
and $P_2=[n_B=0,n_A=-r/(2\varphi_2)]$, that is always outside of the
the physical region. Therefore, the parameter dependence of the system
that differentiates the four classes of behavior is entirely
determined by possible additional solutions given by the intersections
between ${\cal C}_1^{+}$ and ${\cal R}_2$.

The slope of the line ${\cal R}_2$ is $m_{{\cal R}_2}=1/\gamma$, 
independent of $r$, and fixed once we fix the model parameters. 
For large $r$ the intercept of ${\cal R}_2$ tends to
either $\pm \infty$, depending on the sign of $\gamma$. 
This implies that, independent on the value of the
parameters, full consensus on $A$ (the point $P_1$) is the only 
stationary solution for large $r$.
As $r$ decreases, ${\cal R}_2$ translates towards the physical region,
while ${\cal C}_1^{+}$ becomes more and more squeezed
towards its asymptotes $n_A=0$, and $n_A=\varphi n_B-r/(2\varphi_2)+1$
and degenerates into their product as $r\rightarrow 0$. 
In order to understand if and when other solutions appear,
it is useful to analyze first the behavior in the limit of no bias.

\subsubsection*{Limit of no bias ($r\rightarrow0$)}
We now consider the limit of no bias ($r\rightarrow 0$).  In such a
limit ${\cal R}_2$ has equation $n_A=n_B/\gamma-1/\gamma$, and ${\cal
  C}_1$ degenerates into the product of its asymptotes $n_A=0$ and
$n_A= \varphi n_B +1$:
\begin{equation}
\left\{\begin{array}{lr}
n_A (n_A-\varphi n_B-1)=0 
& \,\,\,{\cal C}_1\!\!\\
n_B(\gamma  n_A - n_B+1)=0  
& \,\,\, {\cal C}_2\!\!.
\end{array}
\right. .
\label{curvesr0}
\end{equation}

In this case ${\cal C}_2$ and ${\cal C}_1$ have four intersections:
$P_1=[n_B=0,n_A=1]$, $P_2=[n_B=0,n_A=0]$ (intersections of ${\cal
  R}_1$ with the degenerate hyperbola), and $Q_1=[n_B=1,n_A=0]$,
$Q_2=[n_B^*,n_A^*]$ (intersections of ${\cal R}_2$ with the degenerate
hyperbola), where
\begin{equation} 
\begin{array}{l}
n_A^*=(1+\varphi)/(1-\varphi\gamma)\\
n_B^*=(1+\gamma)/(1-\varphi\gamma)\,.
\end{array}
\end{equation}
$Q_2$ is the only fixed point that depends on the model parameters: it
falls inside the physical region for $\varphi_1\leq 0$, $\gamma_1\leq
0$, and outside in all other cases.  This can be proven by noting that
the conditions $\varphi_1\leq0$, $\gamma_1\leq0$ translate into
$\varphi \leq -1$, and $\gamma\leq-1$, which also give $\varphi \gamma
\geq 1$.  Therefore $(1+\varphi)$, $(1+\gamma)$ and
$(1-\varphi\gamma)$ are all negative, implying $n_A^*\geq0$,
$n_B^*\geq 0$.  Moreover
$n_A^*+n_B^*=(2+\varphi+\gamma)/(1-\varphi\gamma)=1+(1+\varphi+\gamma+\varphi\gamma)/(1-
\varphi\gamma)=1+(1+\varphi)(1+\gamma)/(1-\varphi\gamma)\leq1$.  In
the region $\varphi_1> 0$, $\gamma_1<0$, $n_A^*$ and $n_B^*$ have
opposite sign, therefore one of the two has to be negative, and the
point $Q_2$ falls outside the physical region.  The same reasoning
holds for $\varphi_1< 0$, $\gamma_1>0$.  Stability analysis trivially
gives that the fixed point $P_2$ (with eigenvalues $2\varphi_2$ and
$2\gamma_2$) is always repulsive.  The fixed point $P_1$ (with
eigenvalues $2\gamma_1$ and $-2\varphi_2$) is attractive for
$\gamma_1<0$ and a saddle point for $\gamma_1>0$, while the fixed
point $Q_1$ (with eigenvalues $-2\gamma_2$ and $2\varphi_1$) is
attractive for $\varphi_1<0$ and a saddle point for $\varphi_1>0$.
When physically relevant (for $\varphi_1\leq 0$, $\gamma_1\leq 0$),
the fixed point $Q_2$ is always a saddle point. These results are
summarized in the first column of Fig.~\ref{PD}.

\subsubsection*{Case $\varphi_1 \geq 0$}
For $\varphi_1 \geq 0$ the upper branch ${\cal C}_1^+$ of the
hyperbola only crosses the physical region in the point $P_1$,
therefore in that case the status of total consensus on opinion $A$ is
the only fixed point (see Fig.~\ref{PD}, fourth row).  This can be
proven by looking at the slope of the tangent ${\cal T}$ to 
${\cal C}_1^+$ in $P_1$. 
From Eq.~(\ref{curves}) the equation of
${\cal T}$ is
\begin{equation}
n_A =\varphi/(1+r/(2\varphi_2)) n_B +1\,\,\,\,\,\,\,\,\,\,\, {\cal T}.
\end{equation}
${\cal C}_1^+$ enters the physical region only when the angular
coefficient $m_{{\cal T}}=\varphi/(1+r/(2\varphi_2))$ of ${\cal T}$ is
$m_{{\cal T}}<-1$, but this condition is never met for $\varphi_1 \geq
0$.  This proves that in the region $\varphi_1 \geq 0$ of the
parameter space corresponding to {\it Class IV} models, no transition
occurs (see the phase diagram shown in Fig.~(\ref{fig2:PhaseDiagram}(a))).
In what follows we therefore restrict our analysis to the case
$\varphi_1 < 0$ ($\varphi<-1$).

\subsubsection*{Saddle--node bifurcation line}
For $\varphi_1<0$ it is always $m_{{\cal T}}<-1$, therefore ${\cal
  C}_1^+$ goes through the physical region.  
We want to determine under what conditions a saddle--node bifurcation
occurs, i.e. two additional physical solutions (interceptions
of ${{\cal C}_1}$ and ${{\cal R}_2}$) appear for a critical
value of the bias $r=r_c$.
In general when a
straight line crosses an hyperbola, the two intersections either
lie on the same branch or one on each branch, depending on the slope
of the straight line relative to the slope of the asymptotes.  In our
case, for any $r$ the slopes of the hyperbola asymptotes are $0$ and
$\varphi$, while the slope of ${{\cal R}_2}$ is $1/\gamma$.  Therefore
we must distinguish the two following cases: \\$(a)$ For $\varphi <
1/\gamma <0$ the line ${\cal R}_2$ either intercepts the same branch
of the hyperbola in two points $Q_1$ and $Q_2$ (possibly coincident),
or it has no intersections at all.  \\$(b)$ For $1/\gamma <\varphi$ or
$1/\gamma >0$ the line ${\cal R}_2$ always intercepts the hyperbola in
two points, one on each branch.

The equation for the intercepts is obtained from  Eq.~(\ref{curves2}):
\begin{equation}
\Gamma n_A^2-n_A B(r) + r/(2\varphi_2)=0
\label{doubleintercept}
\end{equation}
where $B(r)=r/(2\varphi_2)(1+\varphi\varphi_2 / \gamma_2)-(1+\varphi)$
and $\Gamma=\varphi \gamma-1$.  If $\Gamma<0$ (case $(b)$) the
discriminant $\Delta= B(r)^2-4\Gamma r/(2\varphi_2)$ is positive for
any value of $r$. In this case there is no saddle--node bifurcation.
If instead $\Gamma>0$ (case $(a)$) $\Delta$ is
positive (i.e. there are two intersections) for $r \le r_c$ given by
the equation
\begin{equation}
r_c^2 A^2-4r_c \varphi_2(A(1+\varphi)+2\Gamma)+4 \varphi_2^2(1+\varphi)^2=0
\label{rc}
\end{equation}
where $A=1+\varphi\varphi_2/\gamma_2$.

At $r=r_c$, the two solutions $Q_1$ and $Q_2$ of
Eq.~(\ref{doubleintercept}) coincide in
$n_A=B(r_c)/2\Gamma=\sqrt{r_c/(2\varphi_2\Gamma)}$. 
For the saddle--node bifurcation to have a physical relevance
the two coincident solutions
$Q_1= Q_2$ must appear inside the physical region.
Requiring $n_A<1$ implies $r_c/(2\varphi_2)<\Gamma$.  Solving
Eq.~(\ref{rc}) for $r_c$ and replacing its value in the previous
inequality gives, after some algebra, the condition
$\gamma_1<\gamma_1^*$, where
\begin{equation}
\gamma_1^*=\varphi_1\gamma_2/(\varphi_1-\varphi_2\gamma_2)\, .
\label{gammastar}
\end{equation}
Therefore a discontinuous physical transition always occurs for
$\varphi_1<0$ and $\gamma_1<\gamma_1^*$.  For $\gamma_1>\gamma_1^*$
the two intersections appear instead outside the physical region.  The
saddle--node bifurcation line is shown in Fig.~(\ref{fig2:PhaseDiagram}(b)), 
where $r_c$ is
plotted versus $\gamma_1$ for fixed negative $\varphi_1$.
As discussed in the folowing, when $r$ is further lowered, the two
intersections might exit the physical region.

\subsubsection*{Transcritical bifurcation line}

From the previous analysis of the case $r\rightarrow 0$ we see that,
in the case $\varphi_1<0$, $\gamma_1<0$, for any value of $r<r_c$ down to
$0$ there are two attractive physical fixed points, $P_1=[0,1]$ 
(consensus on opinion $A$) and $Q_1$ (pluralism), separated by a
saddle-node $Q_2$ also inside the physical region 
(Fig.~\ref{PD}, first row).
We denote models in this parameter region as belonging to 
{\em Class I} or ``Finite Critical Mass" models.

In the region $\varphi_1<0$, $\gamma_1>0$ instead, we find that for
vanishing $r$ the system is always driven to the attractive point
$Q_1=[1,0]$ (consensus on opinion $B$), and that the point $Q_2$
always falls outside the physical region.  
However, very different behaviors occur for finite $r$ depending 
on $\gamma_1$ being below or above the value $\gamma_1^*$: 
\\$(1)$ When $0<\gamma_1<\gamma_1^*$,
two fixed points $Q_1 = Q_2$ appear in the physical region at $r=r_c$
through a saddle--node bifurcation, as discussed above.  Lowering $r$,
$Q_1$ moves to the right (high $n_B$), and $Q_2$ moves to the left
(low $n_B$). In contrast to what happens for $\gamma_1<0$, at some
finite value $r=r^*$ the fixed point $Q_2$ collides with $P_1=[0,1]$
and then exits the physical region (Fig.~\ref{PD}, second row).
When $P_1$ and $Q_2$ collide,
they exchange stability: for $r^*<r<r_c$ $P_1$ and $Q_1$ are both
stable, and $Q_2$ is a saddle point; as $Q_2$ crosses $P_1$ and exits
the physical region, $P_1$ becomes unstable, leaving $Q_1$ as the only
stable point in the physical region.  
Therefore in this case ({\it Class II} models),
lowering $r$, an unstable fixed point exits the physical region at
$r=r^*$.  The condition for $r^*$ can be obtained by imposing that
${\cal R}_2$ in Eq.~(\ref{lines}) goes through $P_1$ for $r=r^*$, yielding
\begin{equation}
r^*=2\gamma_1 \,.
\end{equation}

\noindent\\$(2)$ When $\gamma_1>\gamma_1^*$ instead it is the point $Q_1$
that enters the physical region when lowering $r$: 
this happens either because two coincident intercepts arise
outside the physical region and then one moves inside ($\Gamma>0$), or
because only one intercept with the upper branch exists ($\Gamma<0$),
and enters the physical region at some $r>0$.  In either case, $Q_1$
collides with $P_1$ at some $r=r^*$.  When $Q_1$ and $P_1$ collide
they exchange stability (transcritical bifurcation).  In this case,
for $r>r^*$, $Q_1$ is outside the physical region and the fixed point 
$P_1$ is stable (Fig.~\ref{PD}, third row).
For $r<r^*$ $Q_1$ (pluralism) enters the physical region and becomes stable, 
while $P_1$ (consensus on opinion $A$) becomes unstable. 
There is only one stable fixed point for any value of $r$.
In this case ({\it Class III } models), the
transition occurring at $r^*$ is continuous.  
The transcritical bifurcation line is determined as before $r^*=2\gamma_1$
for all $\gamma_1>0$, and is
always below the saddle--node bifurcation line, as shown in 
Fig.~(\ref{fig2:PhaseDiagram}(b)).

Exactly at $\gamma_1=\gamma_1^*$, the two bifurcations coincide:
$r_c=r^*$, the double intersection appears exactly when $Q_1= Q_2 =P_1$, 
therefore $Q_2$ immediately leaves the physical region as $r$ is
lowered below $r^*$.  The transition is continuous.

\subsubsection*{Special cases}
\paragraph{$\gamma_2=0:$}
In this case, from Eqs.~(\ref{curves}) it turns out that the straight
line ${\cal R}_2$ is horizontal: $n_A = r/(2 \gamma_1)$.  From this
expression it is clear that for $\gamma_1<0$ there is no stationary
solution other than $P_1$ (consensus on $A$) in the physical region.
Hence no transition occurs for models in {\em Class I}.  The limit
$\gamma_2 \to 0$ of Eq.~(\ref{gammastar}) yields $\gamma_1^*=0$
implying that the parameter space corresponding to {\em Class II}
shrinks to zero.  We conclude that for $\gamma_2=0$ no transition
occurs, apart the case $\varphi_1<0$ and $\gamma_1>0$ ({\em Class
  III}), for which the usual continuous transition takes place.

\paragraph{$\varphi_2=0:$}
In this case, the equation for the hyperbola ${\cal C}_1$ becomes
$n_A(-\varphi_1 n_B+r/2)=r/2$. One of the asymptotes of ${\cal C}_1$
becomes vertical, but nothing unusual happens, and the general
analysis still holds.

\bibliography{CC15}

\begin{thebibliography}{31}
\bibitem[{not({\natexlab{a}})}]{nota2}
\bibinfo{note}{We actually simulate a discrete--time dynamics. Since we are
  interested in stationary properties of the system, this is perfectly
  equivalent to simulating the continuous dynamics considered above.}
\bibitem[{not({\natexlab{b}})}]{nota1}
\bibinfo{note}{The probability of $U$ turning into $B$ in a single $B-U$
  interaction is small compared to the probability of $U$ turning into $A$ in a
  single $A-U$ interaction. However, for suitable initial conditions, $B-U$
  interactions occurs more frequently, so that, on average, they compensate the
  advantage that the $A$ state has both in $A-B$ interactions, and due to the
  media bias.}
\end{thebibliography}
\end{document}